\documentclass[10pt,twocolumn,letterpaper]{article}

\usepackage[pagenumbers]{cvpr} 

\patchcmd{\abstract}{\vspace*{12pt}}{\vspace*{4pt}}{}{}
\AtBeginDocument{%
  \raggedbottom
  \AddToHookNext{shipout/after}{\flushbottom}%
}

\definecolor{cvprblue}{rgb}{0.21,0.49,0.74}
\usepackage[pagebackref,breaklinks,colorlinks,allcolors=cvprblue]{hyperref}

\usepackage{graphicx}
\usepackage{multirow}
\usepackage[table]{xcolor}

\definecolor{runnerbg}{RGB}{255,214,180}
\definecolor{bestbg}{RGB}{250,159,204}

\def\confName{CVPR}
\def\confYear{2026}

\newtheorem{proposition}{Proposition}[section]

\newtheorem{remark}{Remark}[section]

\title{Don’t Guess Pixels. Learn Relations! Complex Operators for MRI Reconstruction}

\author{%
Chun-Wun Cheng$^{1}$ \quad Rui Sun$^{1}$ \quad Shujun Wang$^{2}$\\
Peiyuan Jing$^{3}$ \quad Guang Yang$^{3}$ \quad Carola-Bibiane Sch\"onlieb$^{1}$\\
Javier A. Montoya-Zegarra$^{4,5,6,\dagger}$ \quad Angelica I. Aviles-Rivero$^{7,\dagger,*}$\\[0.6em]
{\small $^{1}$University of Cambridge}\\
{\small $^{2}$Hong Kong Polytechnic University}
{\small $^{3}$Imperial College London}\\
{\small $^{4}$ZHAW -- Z\"urcher Hochschule f\"ur Angewandte Wissenschaften}\\
{\small $^{5}$Lucerne University Teaching and Research Hospital}
{\small $^{6}$Universidad Peruana Cayetano Heredia}\\
{\small $^{7}$ YMSC, Tsinghua University}\\[0.4em]
{\footnotesize $^{\dagger}$Equal senior contribution. \quad $^{*}$Corresponding author.}
}

\begin{document}
\addtocontents{toc}{\protect\setcounter{tocdepth}{-1}}
\maketitle
\begin{abstract}
Accelerated MRI requires resolving severe aliasing ambiguities induced by undersampling, which entangles signals from spatially separated locations. Existing learned reconstruction methods combine image priors with data consistency, modeling these acquisition-induced, non-local couplings only implicitly. Instead, we introduce the Encoding-Ambiguity Difference Operator (EADO), which explicitly learns complex-valued pairwise differences across acquisition-coupled locations. EADO constructs a sparse graph integrating local neighborhoods with long-range aliasing offsets from the sampling point-spread function (PSF), modulated by coil sensitivities. A shared phase-covariant predictor estimates pairwise complex differences and nonnegative confidence weights on this graph while preserving interaction phase. At each reconstruction stage, a closed-form quadratic solver reconciles these predictions with measured $k$-space to recover a complex image. Theoretically, we bound the truncation error of discarded interactions, establish gauge covariance, and characterize joint identifiability under a null-space condition. EADO surpasses LMO by $2.96$~dB in PSNR on single-coil IXI ($4\times$ random undersampling) and outperforms LMO-SENSE in SSIM across all twelve unseen-mask settings on Stanford2D.
\end{abstract}    
\section{Introduction}
\label{sec:intro}

\begin{figure}[t]
    \centering
    \includegraphics[width=\columnwidth]{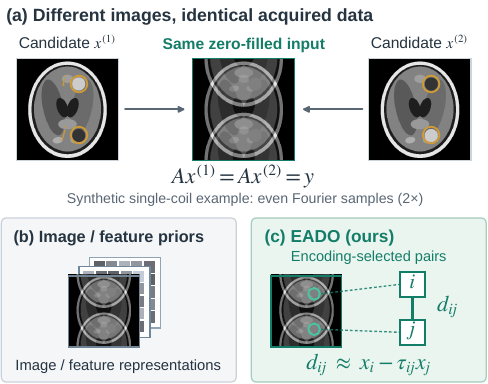}
    \caption{\textbf{Motivation for EADO.} (a) With unit coil sensitivity and single-coil $2\times$ sampling at even Fourier indices along one encoding direction, two synthetic images differ at locations half a field of view apart yet share the acquired data and zero-filled input. (b,c) Schematic comparison of image/feature representations and EADO's explicit prediction of transported complex differences between encoding-selected pixels.}
    \label{fig:eado_motivation}
\end{figure}

Accelerated magnetic resonance imaging (MRI) seeks to shorten acquisition while preserving the anatomical detail needed for image reconstruction. Acquiring fewer k-space measurements reduces scanning time, but missing Fourier information introduces aliasing and makes reconstruction an ill-posed computational imaging problem~\cite{Zbontar2018fastMRI}. Matching the acquired measurements is not sufficient: undersampling can leave distinct images consistent with the same data, as illustrated in Fig.~\ref{fig:eado_motivation}(a). For example, regular undersampling mixes spatially separated structures, so resolving a local artifact can require information from distant image locations. This motivates reconstruction methods that combine learned image priors with acquisition physics~\cite{Sriram2020E2EVarNet}.

Existing methods address this challenge through increasingly expressive priors and reconstruction architectures. Classical model-based approaches exploit coil sensitivity encoding~\cite{Pruessmann1999SENSE}, transform sparsity~\cite{Lustig2007SparseMRI}, and structured low-rank relations~\cite{Haldar2014LORAKS}. These methods rely on predefined models, with sparse and low-rank formulations typically requiring iterative optimization. In contrast, purely data-driven deep learning models—spanning CNN, U-Net, and Transformer architectures—learn direct mappings from aliased inputs to clean images~\cite{Zbontar2018fastMRI,Liang_2021_ICCV}. To bridge physics and learning, unrolled and cascaded networks combine learned regularization with explicit acquisition models~\cite{Hammernik2018Variational,Schlemper2018DeepCascade}: MoDL alternates image denoising with data consistency, while H-DSLR integrates image priors with multichannel Fourier relations~\cite{Aggarwal2019MoDL,Pramanik2020DeepSLR}. More recently, LMO learns local and global feature interactions through neural operators and Mamba-style scanning~\cite{Li_2025_CVPR}. Yet enforcing data consistency alone does not specify which spatial relationships the network should learn. In particular, an image-domain denoiser or a feature-space integral operator does not directly predict complex differences between pixel pairs coupled by the acquisition operator (Fig.~\ref{fig:eado_motivation}(b,c)).

The encoding normal operator describes how forward acquisition and backprojection couple image locations, providing a principled way to select such pixel pairs. For a Cartesian mask shared across coils, each spatial coupling is the product of a sampling-dependent point-spread-function (PSF) coefficient and a coil-sensitivity interaction. Thus, undersampling defines a structured pattern of spatial interactions whose locations, strengths, and complex phases depend directly on the acquisition. This motivates predicting complex differences between selected pixels after accounting for the phase carried by their encoding coupling. These transported differences represent local image structure and nonlocal aliasing relationships in a unified form: the network learns how coupled pixels relate, while an explicit numerical solve reconciles these predictions with the measured data.

We introduce the \emph{Encoding-Ambiguity Difference Operator} (EADO), a reconstruction framework that learns transported complex image differences. EADO first constructs a sparse ambiguity graph by combining local neighbors with aliasing offsets selected from the sampling PSF, where PSF coefficients and coil-sensitivity interactions jointly determine edge transports and strengths. At each stage, a shared predictor estimates complex edge differences and nonnegative integration weights. It combines phase-invariant context with complex features designed to transform consistently under joint image and sensitivity-map phase changes. Because predicted edge relationships need not be mutually consistent, a measurement-anchored quadratic solver reconciles them with acquired $k$-space data, and all stages are trained end-to-end under reconstruction supervision. Unlike MoDL's image-denoising targets or LMO's feature integration, EADO makes acquisition-selected complex relationships the direct targets of reconstruction while avoiding dense pixel-to-pixel matrices via sparse graph operations.

Our analysis provides theoretical guarantees for graph sparsification, phase consistency, and conditional recoverability. A PSF-tail bound quantifies the normal-operator interactions discarded by graph sparsification, while gauge covariance ensures consistency under joint phase transformations of the image and sensitivity maps in exact arithmetic. Under a joint null-space condition, we establish identifiability and derive an error bound for the exact unanchored quadratic formulation, quantifying the effects of measurement noise, edge-target error, and conditioning. Empirically, experiments on single-coil IXI and multicoil Stanford2D validate EADO across diverse sampling trajectories and acceleration factors. On IXI under $4\times$ random undersampling, EADO reaches $42.10$~dB in PSNR, outperforming the reported LMO result by $2.96$~dB. On the Stanford2D validation partition, EADO surpasses LMO-SENSE in SSIM across all twelve unseen-mask configurations spanning three sampling patterns and four acceleration factors. Controlled ablations further confirm the effectiveness of PSF-guided edge selection and learned integration weights.

Our main contributions are summarized as follows:
\begin{itemize}
    \item We introduce an encoding-driven relational formulation of MRI reconstruction that learns transported complex differences between acquisition-coupled pixels rather than direct image corrections. EADO exploits the encoding normal operator to construct a sparse graph of local and PSF-selected aliasing interactions, with phase-aware complex transports.

    \item We develop a phase-covariant difference predictor with adaptive edge weights and a measurement-conditioned quadratic solver. We establish a PSF-tail bound for graph sparsification, prove gauge covariance, and derive a conditional joint-identifiability criterion and error bound linking measurement noise, difference-prediction error, and conditioning.

    \item We demonstrate EADO's effectiveness on single-coil IXI and multicoil Stanford2D, achieving a 2.96 dB PSNR gain over LMO on IXI and higher SSIM across all twelve unseen-mask settings on Stanford2D. Ablations confirm the importance of PSF-guided alias selection and adaptive edge weighting.
\end{itemize}

\begin{figure*}[!t]
\centering
\includegraphics[width=\textwidth]{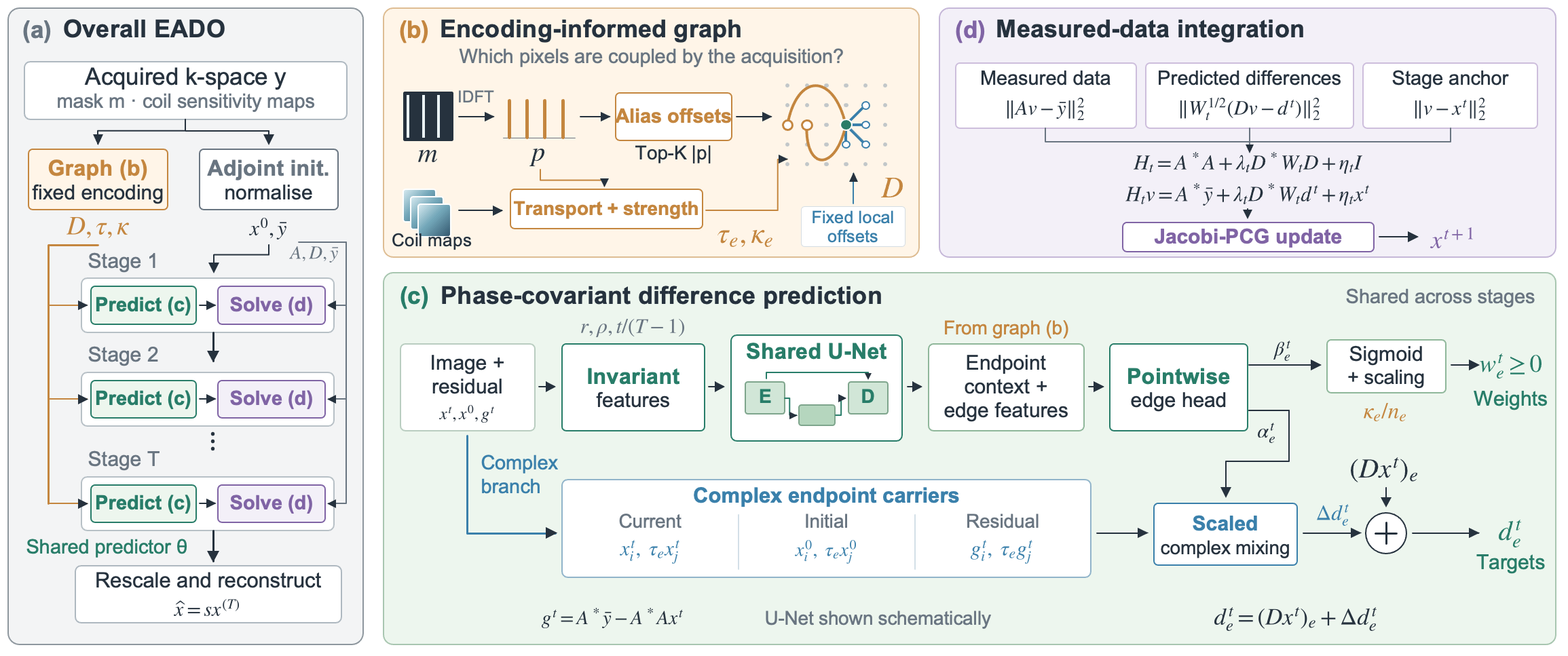}
\caption{Overall architecture of EADO. (a) Reconstruction proceeds through repeated difference-prediction and quadratic-integration stages with a shared predictor. (b) Fixed local connections are augmented by PSF-selected alias connections, with sampling and coil information determining edge transports and strengths. (c) Phase-invariant context controls complex-carrier mixing to predict transported difference targets and nonnegative integration weights. (d) An anchored, measurement-conditioned quadratic system is approximately solved using Jacobi-preconditioned conjugate gradients to update the image.}
\label{fig:eado_overview}
\end{figure*}

\section{Related Work}
\label{sec:related_work}

\textbf{$\blacktriangleright$ Deep Learning for Accelerated MRI Reconstruction.} Learning-based MRI reconstruction ranges from direct image prediction to networks derived from iterative optimization. The fastMRI U-Net baseline~\cite{Zbontar2018fastMRI} learns an image-domain reconstruction mapping, while model-based networks incorporate the acquisition operator into trainable reconstruction stages. Variational networks~\cite{Hammernik2018Variational} learn filters, nonlinearities, and data-fidelity weights within an unrolled gradient-descent scheme. MoDL~\cite{Aggarwal2019MoDL} combines a shared convolutional prior with conjugate-gradient data consistency, while H-DSLR~\cite{Pramanik2020DeepSLR} learns complementary k-space and image-domain priors from structured low-rank reconstruction. E2E-VarNet~\cite{Sriram2020E2EVarNet} jointly learns coil-sensitivity estimation and reconstruction through cascaded updates. These unrolled methods retain the forward model while learning how images should be regularized.

Recent approaches also address sampling degradation and long-range dependencies. PDAC~\cite{Wang2024PDAC} decomposes severe undersampling into a sequence of milder reconstruction problems and learns the intermediate sampling masks. AeSPa~\cite{Joo2025AeSPa} combines attention-guided k-space constraints with scan-specific self-supervision to reconstruct without fully sampled training references. LMO~\cite{Li_2025_CVPR} combines local convolutional integration with global Mamba-style scanning in a neural-operator formulation. These approaches connect learned representations to acquisition constraints or extend the range of feature interactions. EADO takes a complementary direction by explicitly representing relationships between pixels coupled by the encoding operator within a physics-guided reconstruction framework.

\textbf{$\blacktriangleright$ Graph-Based and Nonlocal Image Priors}
Nonlocal image priors exploit recurring structures by relating spatially separated pixels or patches. Yang and Jacob~\cite{Yang2013Nonlocal} formulate nonlocal regularization through robust distances between patches, while PANO~\cite{Qu2014PANO} constructs sparse representations of similar patches for MRI reconstruction with data fidelity. DeepGLR~\cite{Zeng_2019_CVPR_Workshops} constructs a graph from learned features and restores images through a differentiable quadratic optimization layer with a graph-Laplacian regularizer. Valsesia et al.~\cite{Valsesia2020DeepGraph} incorporate nonlocal processing into graph-convolutional denoisers by computing neighborhoods from hidden-feature similarities.

EADO makes acquisition-selected complex differences the learned targets of reconstruction. Its nonlocal connections follow the sampling point-spread function, while complex transports reflect both sampling and coil-sensitivity interactions. Whereas standard graph-Laplacian regularization favors small differences, EADO learns how encoding-coupled pixels should differ after phase transport. A phase-covariant predictor estimates these generally nonzero targets and their nonnegative integration weights. Because the predicted relations need not be mutually compatible, a quadratic reconstruction stage reconciles them with measured k-space to recover a single complex image.

\section{Methodology}
\label{sec:method}

EADO reconstructs undersampled MRI by learning complex differences across encoding-selected pixel pairs rather than directly predicting image corrections. As shown in Fig.~\ref{fig:eado_overview}, each stage uses a shared phase-covariant predictor to estimate differences and weights on the encoding-informed graph, then integrates them with measured k-space through a quadratic solve. We establish three properties of this formulation: PSF-tail control, gauge covariance, and joint identifiability.

\subsection{Problem Setting}
\label{sec:formulation}

We consider accelerated MRI reconstruction from undersampled multi-coil measurements:
\begin{equation}
y=Ax^\star+\varepsilon,\qquad (Ax)_c=M\mathcal F(S_cx),
\label{eq:forward}
\end{equation}
where $x^\star\in\mathbb C^N$ ($N=HW$) is the target image on a periodic grid, $S_c=\operatorname{diag}(s_c)$ denotes coil sensitivities, $M=\operatorname{diag}(m)$ is the shared binary Cartesian sampling mask, and $\mathcal F$ is the centered orthonormal Fourier transform. Measurements and noise are zero outside the acquired support; $C=1$ and $s_1\equiv1$ give single-coil MRI. Our goal is to resolve encoding-induced ambiguity through learned complex edge relationships while preserving measured-data fidelity.

\subsection{Encoding-Induced Ambiguity Graph}
\label{sec:graph}

The encoding normal operator reveals sampling-dependent pixel couplings:
\begin{equation}
\begin{aligned}
(A^*A)_{ij}&=p(i-j)c_{ij},\\
c_{ij}&=\sum_{c=1}^{C}\overline{s_{c,i}}s_{c,j},\\
p&=\operatorname{IDFT}\bigl(\operatorname{ifftshift}(m)\bigr).
\end{aligned}
\label{eq:psf}
\end{equation}
The IDFT uses conventional $1/N$ normalization, so $p(0)=\rho=N^{-1}\sum_km_k$. For unit-sensitivity single-coil MRI, $\dim\ker A=(1-\rho)N$; multiple coils may reduce this ambiguity. The PSF identifies candidate interactions, while $c_{ij}$ captures coil-dependent coupling.

We combine four local offsets $\Delta_{\rm loc}=\{(1,0),(0,1),(1,1),(1,-1)\}$ with $K$ alias offsets selected by the largest eligible $|p(\delta)|$. We exclude zero and both orientations of local offsets, retain one representative of each remaining class $\{\delta,-\delta\}$, and resolve ties by flattened-index order. Each offset defines edges $e=(i,j)$ with $j=i-\delta$ under periodic indexing. Selected offset classes are assumed distinct on the grid.

\begin{proposition}[PSF-tail certificate]
\label{prop:psftail}
Let $\mathcal O$ contain zero and both orientations of every retained offset class, including local offsets. Define $\widetilde N_{ij}=\mathbf 1_{\{i-j\in\mathcal O\}}p(i-j)c_{ij}$. Then
\begin{equation}
\|A^*A-\widetilde N\|_{\mathrm{op}}
\le
\left(\sum_{\delta\notin\mathcal O}|p(\delta)|\right)
\sum_{c=1}^{C}\|s_c\|_\infty^2.
\label{eq:tailbound}
\end{equation}
\end{proposition}

The certificate quantifies omitted normal-operator interactions; it does not assert $D^*WD=A^*A$. Since $m$ is real, $|p(-\delta)|=|p(\delta)|$. Ranking by $|p(\delta)|$ minimizes the discarded PSF $\ell_1$ tail at fixed $K$ when all eligible classes contain two distinct offsets. In the presence of self-inverse classes, exact tail-optimal selection instead ranks by $|\{\delta,-\delta\}|\,|p(\delta)|$. This distinction does not change the implemented magnitude-based ranking. The proof is in Appendix.

To preserve complex encoding interactions, let $r_i=\sum_{c=1}^{C}|s_{c,i}|^2$ denote sensitivity power, $\zeta_e\in\{0,1\}$ indicate an alias edge, and $\epsilon_g=10^{-12}$. Define
\begin{equation}
\begin{aligned}
q_e&=c_{ij}\big[(1-\zeta_e)+\zeta_ep(\delta)\big],\\
\tau_e&=\frac{q_e}{\max(|q_e|,\epsilon_g)},\\
\chi_{ij}&=\min\!\left(1,\frac{|c_{ij}|}{\max(\sqrt{r_ir_j},\epsilon_g)}\right),\\
\omega_\delta&=\min\!\left(\frac{|p(\delta)|}{\max(\rho,\epsilon_g)},1\right),\\
\kappa_e&=\chi_{ij}\big[(1-\zeta_e)+\zeta_e\omega_\delta\big].
\end{aligned}
\label{eq:transport}
\end{equation}
The transport $\tau_e$ carries the interaction phase when $|q_e|\ge\epsilon_g$ and always satisfies $|\tau_e|\le1$, while $\kappa_e$ controls its strength. The resulting complex difference operator is
\begin{equation}
(Dx)_e=x_i-\tau_ex_j.
\label{eq:difference}
\end{equation}
Local edges capture neighborhood relationships, while alias edges expose nonlocal encoding couplings. Both $D$ and its Hermitian adjoint $D^*$ are applied without constructing dense matrices.

\subsection{Phase-Covariant Difference Prediction}
\label{sec:predictor}

We initialize with a sensitivity-power-compensated adjoint reconstruction:
\begin{equation}
\begin{aligned}
z_i&=\frac{(A^*y)_i}{\max(r_i,10^{-8})},\\
s&=\max\!\left(\sqrt{N^{-1}\sum_i|z_i|^2},10^{-6}\right),\\
x^0&=z/s,\qquad \bar y=y/s.
\end{aligned}
\label{eq:normalization}
\end{equation}
The positive scale $s$ is treated as constant during differentiation; the final image is $\widehat x=sx^{(T)}$.

At stage $t\in\{0,\ldots,T-1\}$, define the measured-data residual
\begin{equation}
b=A^*\bar y,\qquad g^t=b-A^*Ax^t.
\label{eq:residual}
\end{equation}
A shared U-Net extracts phase-invariant scalar context from $|x^t|$, $|x^0|$, $|g^t|$, the real and imaginary parts of $\overline{x^t}g^t/(1+|x^t||g^t|)$, $\log(1+r)$, $\rho$, and $t/(T-1)$. Image-valued products and divisions in these features are elementwise.

For each edge $e=(i,j)$, a shared pointwise head combines endpoint context with $\kappa_e$, $\zeta_e$, a sinusoidal offset embedding, and the real and imaginary parts of $\overline{x_i^t}\tau_ex_j^t/(1+|x_i^t||\tau_ex_j^t|)$. It predicts a real logit $\beta_e^t$ and six complex coefficients $\alpha_{e\ell}^t$, with real and imaginary parts separately bounded by $\tanh$. Because the head receives only gauge-invariant features, its predicted coefficients and weights are gauge-invariant.

Using six complex carriers
\begin{equation}
h_e^t=[x_i^t,\tau_ex_j^t,x_i^0,\tau_ex_j^0,g_i^t,\tau_eg_j^t],
\label{eq:carriers}
\end{equation}
we predict an edge target and nonnegative integration weight:
\begin{equation}
\begin{aligned}
d_e^t&=(Dx^t)_e+\frac{0.25}{\sqrt6}\sum_{\ell=1}^{6}\alpha_{e\ell}^th_{e\ell}^t,\\
w_e^t&=\frac{\kappa_e}{n_e}\bigl(0.05+1.95\sigma(\beta_e^t)\bigr),
\end{aligned}
\label{eq:target}
\end{equation}
where $\sigma$ is the logistic sigmoid and $n_e=4$ for local edges or $K$ for alias edges ($K\ge1$). The complex carriers retain phase information, while the invariant coefficients determine image-dependent corrections to the current transported differences.

\begin{proposition}[Gauge covariance]
\label{prop:gauge}
Let $U=\operatorname{diag}(u)$ with $|u_i|=1$, and transform $x\mapsto Ux$ and $s_{c,i}\mapsto\overline{u_i}s_{c,i}$. Define $(U_{\mathcal E}z)_e=u_i z_e$ for each edge $e=(i,j)$. Then $A'=AU^*$, $\tau_{ij}'=u_i\overline{u_j}\tau_{ij}$, and $D'U=U_{\mathcal E}D$. The predictor's scalar inputs, coefficients, and weights remain invariant, while $d^{t\prime}=U_{\mathcal E}d^t$. The exact quadratic stage and Jacobi-preconditioned PCG iterations are covariant in exact arithmetic under a covariant initialization and stopping rule. Consequently, $\widehat x'=U\widehat x$ and $|\widehat x'|=|\widehat x|$.
\end{proposition}

In particular, $H_t'=UH_tU^*$ and $\operatorname{diag}(H_t')=\operatorname{diag}(H_t)$, so the Jacobi preconditioner is invariant. The transport normalization preserves covariance even when $|q_e|<\epsilon_g$. The full proof is in Appendix.

\subsection{Quadratic Integration with Measured Data}
\label{sec:solve}

Independently predicted edge differences may conflict around graph cycles. EADO integrates these relationships into a single complex image while fitting acquired measurements. With $W_t=\operatorname{diag}(w^t)$ and predictions fixed within stage $t$, define
\begin{equation}
\begin{split}
J_t(v)={}&\frac12\|Av-\bar y\|_2^2+\frac{\lambda_t}{2}\|W_t^{1/2}(Dv-d^t)\|_2^2\\
&+\frac{\eta_t}{2}\|v-x^t\|_2^2.
\end{split}
\label{eq:energy}
\end{equation}
The terms retain measured-data fidelity, fit the predicted differences, and anchor the update to the current estimate. The unique minimizer satisfies
\begin{equation}
\begin{aligned}
H_tv_t^\star&=b+\lambda_tD^*W_td^t+\eta_tx^t,\\
H_t&=A^*A+\lambda_tD^*W_tD+\eta_tI.
\end{aligned}
\label{eq:system}
\end{equation}
Since $\lambda_t>0$, $W_t\succeq0$, and $\eta_t>0$, we have $H_t\succeq\eta_tI\succ0$. Thus every frozen quadratic has a unique minimizer, with $\|H_t^{-1}\|_{\mathrm{op}}\le1/\eta_t$. Equivalently,
\begin{equation}
v_t^\star-x^t=H_t^{-1}\!\left[g^t+\lambda_tD^*W_t(d^t-Dx^t)\right].
\label{eq:correction}
\end{equation}
This identity exposes EADO's mechanism: the update combines the measured-data residual with learned corrections to encoding-informed complex differences. The proximal anchor guarantees stage well-posedness, not reconstruction accuracy.

We approximate Eq.~\eqref{eq:system} using Jacobi-preconditioned conjugate gradients (PCG), initialized at $x^t$ with $P_t=\operatorname{diag}(H_t)$. The executed iterations define $x^{t+1}$ and are differentiated end to end. Positive definiteness holds for each frozen quadratic, while the finite PCG budget yields an approximate solution with soft rather than exact data fidelity.

\begin{proposition}[Joint identifiability and error bound]
\label{prop:identifiability}
For fixed $D$, nonnegative $W$, and $\lambda>0$, exact measurements $Ax$ and weighted differences $W^{1/2}Dx$ jointly determine $x$ uniquely if and only if
\begin{equation}
\ker A\cap\ker(W^{1/2}D)=\{0\}
\quad\Longleftrightarrow\quad
A^*A+\lambda D^*WD\succ0.
\label{eq:identifiability}
\end{equation}
Suppose this condition holds, $y=Ax^\star+\varepsilon$, and $d=Dx^\star+\delta$. Then the minimizer $\widehat x$ of the unanchored quadratic satisfies
\begin{equation}
\|\widehat x-x^\star\|_2
\le
\frac{\sqrt{\|\varepsilon\|_2^2+\lambda\|W^{1/2}\delta\|_2^2}}
{\sigma_{\min}\!\left(
\begin{bmatrix}
A\\
\sqrt{\lambda}W^{1/2}D
\end{bmatrix}
\right)}.
\label{eq:identerror}
\end{equation}
\end{proposition}

The bound separates measurement noise, difference-prediction error, and joint conditioning. It explains how accurate edge information can constrain ambiguities left by undersampling. However, learned targets are not independent measurements, and the result concerns the exact unanchored problem rather than the anchored finite-PCG reconstruction. The proof is in Appendix.

With $L=4+K$ edge directions per pixel, each PCG matrix action costs $O(CN\log N+LN)$ using FFT-based encoding and sparse graph operations.

\begin{table*}[!t]
\centering
\caption{Quantitative MRI reconstruction at nominal $4\times$ and $8\times$ acceleration. PSNR is reported in dB; $\uparrow$ indicates higher is better. Best results are \textbf{bold} with light blue shading; second-best results are \underline{underlined}. Dashes denote unavailable results.}
\label{tab:mri_legacy_comparison}
\small
\definecolor{mribest}{RGB}{220,235,249}
\renewcommand{\arraystretch}{1.12}
\setlength{\tabcolsep}{3pt}
\begin{tabular*}{\textwidth}{@{\extracolsep{\fill}}cl*{12}{c}@{}}
\toprule
\multirow{3}{*}{AR}
& \multirow{3}{*}{Method}
& \multicolumn{6}{c}{\textbf{IXI (Brain)}}
& \multicolumn{6}{c}{\textbf{Stanford 2D}} \\
\cmidrule(lr){3-8}\cmidrule(lr){9-14}
&
& \multicolumn{2}{c}{Random}
& \multicolumn{2}{c}{Radial}
& \multicolumn{2}{c}{Equispaced}
& \multicolumn{2}{c}{Random}
& \multicolumn{2}{c}{Radial}
& \multicolumn{2}{c}{Equispaced} \\
\cmidrule(lr){3-4}\cmidrule(lr){5-6}\cmidrule(lr){7-8}
\cmidrule(lr){9-10}\cmidrule(lr){11-12}\cmidrule(lr){13-14}
&
& PSNR$\uparrow$ & SSIM$\uparrow$
& PSNR$\uparrow$ & SSIM$\uparrow$
& PSNR$\uparrow$ & SSIM$\uparrow$
& PSNR$\uparrow$ & SSIM$\uparrow$
& PSNR$\uparrow$ & SSIM$\uparrow$
& PSNR$\uparrow$ & SSIM$\uparrow$ \\
\midrule
\multirow{7}{*}{$4\times$}
& U-Net
& 31.28 & 0.954 & 34.03 & 0.935 & 30.22 & 0.946
& 36.72 & 0.926 & 40.00 & \underline{0.944} & 36.63 & \underline{0.926} \\

& SwinIR
& 32.51 & 0.962 & 35.57 & 0.940 & 31.30 & 0.951
& - & - & - & - & - & - \\

& U-Mamba
& 32.10 & 0.958 & 34.03 & 0.930 & 30.92 & 0.945
& 35.81 & 0.916  &39.48  &0.940  &35.79  &0.916  \\

& H-DSLR
& 36.01 & 0.982 & 45.31 & 0.994 & 33.65 & 0.968
& 37.02 & \cellcolor{mribest}\textcolor{black}{\textbf{0.929}} & \cellcolor{mribest}\textcolor{black}{\textbf{41.15}} & \cellcolor{mribest}\textcolor{black}{\textbf{0.950}} & 36.38 & 0.924 \\

& PGIUN
& 37.98 & 0.985
& 47.09 & 0.994
& 35.51 & 0.978
& - & -
& - & -
& - & - \\

& LMO
& \underline{39.14} & \underline{0.986}
& \underline{48.16} & \underline{0.996}
& \underline{35.70} & \underline{0.980}
& \underline{38.18} & \underline{0.927}
& \underline{40.97} & 0.943
& \underline{38.03} & 0.922 \\

\addlinespace[1.5pt]
& \textbf{EADO (Ours)}
& \cellcolor{mribest}\textcolor{black}{\textbf{42.10}} & \cellcolor{mribest}\textcolor{black}{\textbf{0.993}} & \cellcolor{mribest}\textcolor{black}{\textbf{49.72}} & \cellcolor{mribest}\textcolor{black}{\textbf{0.999}} & \cellcolor{mribest}\textcolor{black}{\textbf{37.42}} & \cellcolor{mribest}\textcolor{black}{\textbf{0.985}}
& \cellcolor{mribest}\textcolor{black}{\textbf{38.59}} & 0.925 & 40.93 & 0.943 & \cellcolor{mribest}\textcolor{black}{\textbf{39.14}} & \cellcolor{mribest}\textcolor{black}{\textbf{0.935}} \\
\midrule

\multirow{7}{*}{$8\times$}
& U-Net
& 29.06 & 0.932 & 29.86 & 0.890 & 27.91 & 0.922
& 34.33  & 0.898 & 36.58  &0.912  &33.79  &0.890\\

& SwinIR
& 30.13 & 0.947 & 29.79 & 0.891 & 28.54 & 0.925
& - & - & - & - & - & - \\

& U-Mamba
& 29.88 & 0.935 & 29.31 & 0.879 & 28.06 & 0.923
&  33.39 &0.884  &36.22  &0.908  &32.96  &0.876  \\

& H-DSLR
& 31.96 & 0.955 & 34.50 & 0.971 & 29.80 & 0.938
&  33.35 &0.887  &37.34  &0.920  &32.54  &0.876  \\

& PGIUN
& \underline{34.15} & 0.970
& 36.21 & 0.980
& 31.74 & 0.957
& - & -
& - & -
& - & - \\

& LMO
& \cellcolor{mribest}\textcolor{black}{\textbf{34.26}} & \underline{0.971}
& \underline{36.53} & \underline{0.983}
& \underline{31.80} & \underline{0.960}
& \underline{35.49} & \cellcolor{mribest}\textcolor{black}{\textbf{0.909}}
& \cellcolor{mribest}\textcolor{black}{\textbf{38.39}} & \cellcolor{mribest}\textcolor{black}{\textbf{0.927}}
& \underline{34.87} & \underline{0.897} \\

\addlinespace[1.5pt]
& \textbf{EADO (Ours)}
& 34.02 & \cellcolor{mribest}\textcolor{black}{\textbf{0.973}} & \cellcolor{mribest}\textcolor{black}{\textbf{37.60}} & \cellcolor{mribest}\textcolor{black}{\textbf{0.985}} & \cellcolor{mribest}\textcolor{black}{\textbf{33.68}} & \cellcolor{mribest}\textcolor{black}{\textbf{0.971}}
&  \cellcolor{mribest}\textcolor{black}{\textbf{35.83}} & \underline{0.905}  & \underline{38.16}  &\underline{0.921}  &\cellcolor{mribest}\textcolor{black}{\textbf{35.39}}  & \cellcolor{mribest}\textcolor{black}{\textbf{0.904}} \\
\bottomrule
\end{tabular*}
\end{table*}
\begin{figure*}[!t]
\centering
\includegraphics[width=\textwidth]{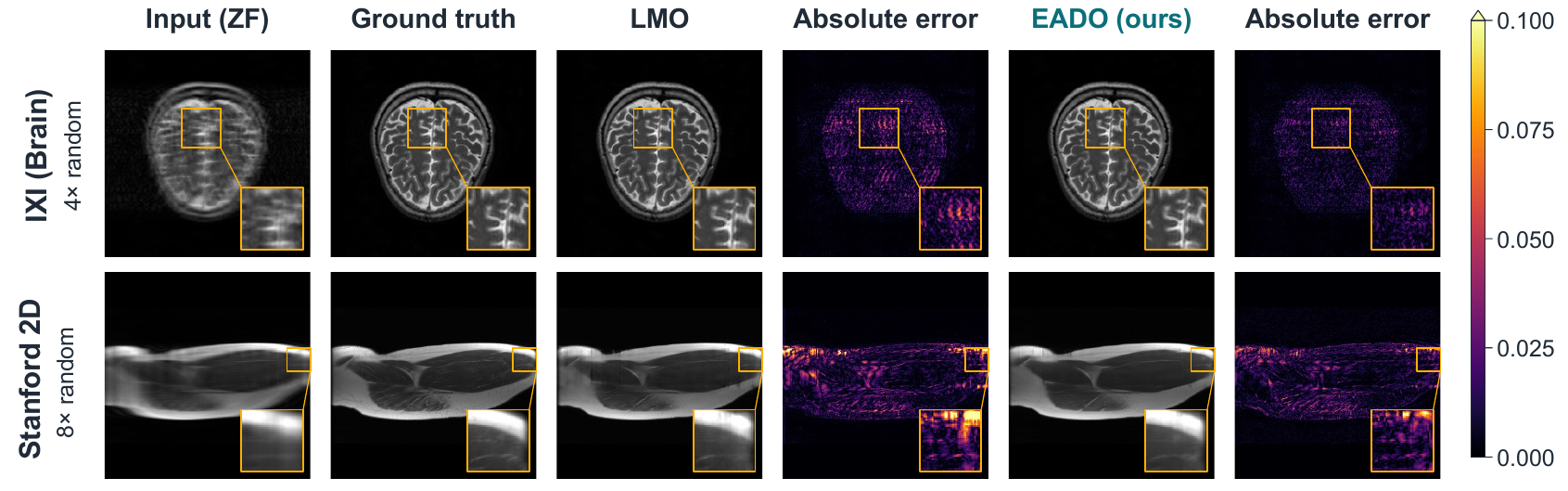}
\caption{Qualitative MRI reconstruction comparisons on IXI brain images with $4\times$ random undersampling (top) and Stanford2D acquisitions with $8\times$ random undersampling (bottom). Columns show the zero-filled input (ZF), ground truth, LMO reconstruction and its absolute error, and EADO reconstruction and its absolute error. Yellow boxes mark regions shown in enlarged insets. Brighter colours in the error maps indicate larger discrepancies from the reference.}
\label{fig:mri_qualitative}
\end{figure*}

\subsection{Training and Inference}
\label{sec:learning}

We train the shared predictor and stage-specific strengths end to end using final-image magnitude supervision. For magnitude target $a$ and output conversion $\mathcal R$, the objective is
\begin{equation}
\mathcal L(\theta)=\mathbb E\!\left[\frac1{|\Omega|}\sum_{i\in\Omega}\sqrt{\bigl(\mathcal R(sx^{(T)})_i-a_i\bigr)^2+\epsilon^2}\right],
\label{eq:loss}
\end{equation}
where $\epsilon=10^{-3}$ and $\Omega$ indexes supervised output pixels. For single-coil IXI, $\mathcal R(x)=|x|$. For multicoil Stanford, $\mathcal R$ returns the center-cropped root-sum-of-squares magnitude $\sqrt{\sum_c|S_cx|^2}$, with $\Omega$ indexed in cropped coordinates. Measurements and targets use the same dataset-specific intensity units, while $s$ provides additional internal normalization.

We use $T=8$ stages, $K=6$ alias offsets, U-Net width $48$, and eight PCG iterations per stage (relative-residual tolerance $10^{-6}$), with $656{,}621$ trainable parameters. The stage-specific strengths are

\begin{equation}
\lambda_t=0.01+0.99\sigma(\ell_t),\qquad
\eta_t=0.005+0.195\sigma(e_t).
\label{eq:strengths}
\end{equation}
At inference, the learned parameters remain fixed while edge targets and weights are recomputed at every stage. Ground-truth images are required only during training.

\newcommand{\Sam}[1]{{\color{red} #1}}
\section{Experiments}
\label{sec:experiments}

We evaluate EADO on single-coil and multicoil MRI reconstruction using IXI dataset and Stanford2D acquisitions, respectively. Experiments cover random, equispaced, and radial masks at nominal acceleration factors of $4\times$ and $8\times$, with reconstruction quality measured by PSNR and SSIM.

\begin{figure*}[h!]
\centering
\includegraphics[width=0.9\textwidth]{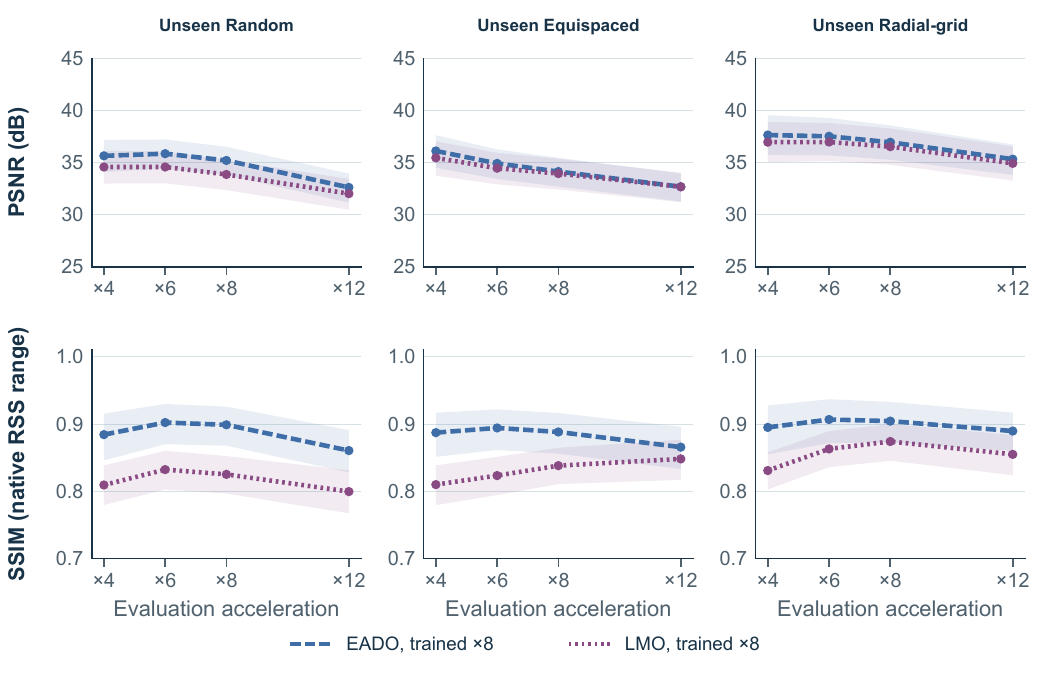}
\caption{Generalisation to unseen masks on the Stanford2D validation partition. EADO and LMO-SENSE models trained at $8\times$ are evaluated with unseen random, equispaced, and radial-grid masks at $4\times$, $6\times$, $8\times$, and $12\times$. All settings retain 15 fixed central ACS columns. The top row reports PSNR; the bottom row reports SSIM using the native RSS range. Blue dashed curves denote EADO, and purple dotted curves denote LMO-SENSE.}
\label{fig:stanford_unseen_masks}
\end{figure*}

\paragraph{Datasets.}
For single coil reconstruction, we use T2-weighted brain images from IXI \footnote{\url{https://brain-development.org/ixi-dataset/}}. We split these subjects into 401 for training, 58 for validation, and 116 for testing, and extract ten fixed $256\times256$ slices per subject, yielding 4,010, 580, and 1,160 slices, respectively. For multicoil reconstruction, we use Stanford2D FSE \footnote{\url{http://mridata.org/list}}, comprising 89 volumes with native complex coil measurements. We use 71 volumes (1,547 slices) for training and 18 volumes (490 slices) for validation. Sensitivity maps are estimated from 15 observed central calibration columns by normalizing the low-resolution coil images with their root-sum-of-squares (RSS) image. 

\paragraph{Competing methods.}
We compare EADO with seven representative methods: the image-restoration networks U-Net~\cite{Zbontar2018fastMRI}, SwinIR~\cite{Liang_2021_ICCV}, and U-Mamba~\cite{Ma2024UMamba}; the unrolled reconstruction methods HQS-Net~\cite{pmlr-v172-xin22a}, H-DSLR~\cite{Pramanik2020DeepSLR}, and PGIUN~\cite{jiang2024pgiun}; and the neural-operator method LMO~\cite{Li_2025_CVPR}. IXI baseline values are taken from the comparisons reported in LMO. On Stanford2D, the LMO baseline is a multicoil adaptation that combines its convolution and scanning components with SENSE-based data consistency. This adaptation and EADO use the same volume split, sampling masks, and RSS evaluation protocol.
\paragraph{Main Results.}
{Table~\ref{tab:mri_legacy_comparison} summarises the quantitative comparisons across three sampling patterns and two acceleration factors. We observe a clear advantage on IXI, where our method achieves the highest SSIM in all six settings and the highest PSNR in five. Specifically, under $4\times$ random sampling, our method reaches $42.10$~dB. This exceeds LMO and PGIUN by $2.96$ and $4.12$~dB, respectively. Under $4\times$ radial sampling, our method achieves $49.72$~dB, compared with $48.16$~dB for LMO. The PSNR gains over LMO persist at $8\times$ for radial and equispaced sampling, reaching $1.07$ and $1.88$~dB, respectively. In the remaining random-sampling case, SSIM is still the highest, while PSNR is $0.24$~dB below LMO. Compared with U-Net, SwinIR, and U-Mamba, our method also achieves higher PSNR and SSIM across all IXI settings. These results suggest that explicitly learning complex differences along encoding-coupled pixels provides an effective reconstruction prior across distinct aliasing structures, rather than relying solely on implicit image or feature representations.}

The multicoil Stanford2D results further demonstrate the applicability of this formulation beyond single-coil reconstruction. EADO achieves the highest PSNR under random and equispaced sampling at both acceleration factors, with its most pronounced joint improvements under equispaced masks. At $4\times$, EADO achieves $39.14$~dB and $0.935$ SSIM, improving upon LMO by $1.11$~dB and $0.013$ SSIM; these advantages persist at $8\times$, with gains of $0.52$~dB and $0.007$ SSIM. The equispaced results are particularly relevant to EADO's design, since structured undersampling induces nonlocal pixel couplings that are explicitly represented through PSF-selected graph edges and complex phase transports. Although performance gains are less uniform under random and radial sampling, the results support the value of incorporating acquisition-induced relationships directly into the learned reconstruction operator.

\textbf{Qualitative comparison.}
Figure~\ref{fig:mri_qualitative} presents visual comparisons on IXI dataset under $4\times$ random undersampling and Stanford2D FSD under $8\times$ random undersampling. The zero-filled inputs exhibit substantial loss of detail, while both learned methods recover clearer anatomical structures. Compared with LMO, EADO more closely follows the ground-truth appearance in the highlighted regions. In the IXI example, the enlarged inset shows better preservation of fine structural detail, accompanied by less prominent residuals in the absolute-error map. The Stanford2D example similarly shows reduced residual error around the highlighted boundary. These examples complement the quantitative comparisons by illustrating EADO's ability to preserve local image structure across single-coil and multicoil reconstruction settings. Additional visualization results can be found in the Appendix.

\paragraph{Generalisation to unseen masks.}
Figure~\ref{fig:stanford_unseen_masks} compares EADO and LMO under unseen sampling masks and varying evaluation accelerations on Stanford2D. Both models are trained at $8\times$, while evaluation covers $4\times$, $6\times$, $8\times$, and $12\times$ with 15 fixed central ACS columns. Our first observation is that EADO achieves higher SSIM in all twelve settings across random, equispaced, and radial-grid sampling. The PSNR advantage is clearest under random sampling and remains visible under radial-grid sampling. For equispaced sampling, the PSNR gap narrows as acceleration increases, and the two curves nearly coincide at $12\times$, although EADO retains an SSIM advantage. At the highest acceleration, both methods exhibit lower PSNR across all three sampling patterns. Under equispaced sampling, EADO achieves substantially higher SSIM than LMO at low and intermediate accelerations, despite closely matched PSNR. Interestingly, LMO exhibits increasing SSIM as acceleration rises, whereas EADO maintains consistently high structural similarity across the evaluated range. Consequently, the SSIM gap narrows at higher accelerations, but EADO retains its advantage throughout. These results suggest that encoding-aware complex difference modeling provides robust structural fidelity under unseen sampling patterns, with benefits that are not fully reflected by PSNR. Overall, these results support EADO's ability to generalise to unseen masks and acceleration factors different from the training rate, with the most consistent advantage appearing in SSIM.

\begin{table}[t]
\centering
\caption{Model complexity and component ablations. (a) Parameter counts and estimated GFLOPs; LMO uses its SENSE adaptation for the GFLOPs comparison. (b) IXI ablations with $8\times$ random undersampling and SSIM data range 2. PSNR drops are relative to the independently retrained full model. Best values are \textbf{bold} with blue shading; second-best values are \underline{underlined}.}
\label{tab:complexity_components}
\footnotesize
\definecolor{complexitybest}{RGB}{220,235,249}
\setlength{\tabcolsep}{2pt}
\renewcommand{\arraystretch}{1.16}
\begin{tabular*}{\columnwidth}{@{\extracolsep{\fill}}lrr@{}}
\toprule
\multicolumn{3}{@{}l}{\textbf{(a) Model complexity}} \\
\addlinespace[2pt]
Model & Parameters $\downarrow$ & \shortstack{Estimated\\GFLOPs $\downarrow$} \\
\midrule
LMO & \underline{2,021,844} & \cellcolor{complexitybest}\textcolor{black}{\textbf{738.15}} \\
\textbf{EADO (Ours)} & \cellcolor{complexitybest}\textcolor{black}{\textbf{656,621}} & \underline{744.99} \\
\bottomrule
\end{tabular*}
\par\vspace{6pt}
\begin{tabular*}{\columnwidth}{@{\extracolsep{\fill}}lrrrr@{}}
\toprule
\multicolumn{5}{@{}l}{\textbf{(b) Controlled component removals}} \\
\addlinespace[2pt]
Variant & \shortstack{PSNR $\uparrow$\\(dB)} & SSIM $\uparrow$ & NRMSE $\downarrow$ & \shortstack{PSNR drop\\(dB)} \\
\midrule
\textbf{Full EADO} & \cellcolor{complexitybest}\textcolor{black}{\textbf{34.0175}} & \cellcolor{complexitybest}\textcolor{black}{\textbf{0.972515}} & \cellcolor{complexitybest}\textcolor{black}{\textbf{0.105320}} & \cellcolor{complexitybest}\textcolor{black}{\textbf{0.0000}} \\
\addlinespace[3pt]
\shortstack[l]{Local connections\\only} & 33.7646 & 0.971078 & 0.108308 & 0.2529 \\
\addlinespace[3pt]
\shortstack[l]{Fixed alias\\offsets} & 33.2107 & 0.967701 & 0.115173 & 0.8068 \\
\addlinespace[3pt]
\shortstack[l]{Fixed edge-weight\\multiplier} & \underline{33.8606} & \underline{0.971629} & \underline{0.107240} & \underline{0.1570} \\
\bottomrule
\end{tabular*}
\end{table}

\paragraph{Model complexity} EADO reduces the learned model size while maintaining an estimated computational cost close to LMO (Table~\ref{tab:complexity_components}(a)). It uses 656,621 parameters, 67.5\% fewer than LMO. With the SENSE adaptation used for the multicoil comparison, LMO requires an estimated 738.15 GFLOPs, compared with 744.99 for EADO, a difference of 0.93\%. This comparison separates the compact shared predictor from the cost of reconstruction, which also includes repeated encoding operations and quadratic solves.

\paragraph{Component ablations.}The full EADO model achieves the highest PSNR and SSIM and the lowest NRMSE among the controlled variants in Table~\ref{tab:complexity_components}(b). On IXI with $8\times$ random undersampling, restricting the graph to local connections reduces PSNR by 0.253\,dB. Replacing PSF-selected alias offsets with fixed offsets causes the largest drop, 0.807\,dB, supporting acquisition-dependent selection of nonlocal connections. Fixing the edge-weight multiplier lowers PSNR by 0.157\,dB, supporting adaptive weighting when integrating predicted differences.

\section{Conclusion}
\label{sec:conclusion}
We presented EADO, an encoding-aware MRI reconstruction framework that learns complex differences between acquisition-coupled pixels. A sparse graph combines local neighbours with PSF-selected alias connections, while a phase-covariant predictor estimates transported differences and adaptive weights. A quadratic integration stage reconciles these predictions with measured $k$-space data. Our analysis bounds the normal-operator interactions discarded by graph sparsification, establishes gauge covariance, and characterises identifiability under a joint null-space condition.  Experiments on single-coil IXI and multicoil Stanford2D validate the effectiveness of learning these explicit relationships, achieving substantial PSNR gains on IXI and higher SSIM than LMO-SENSE across all twelve unseen-mask settings on the Stanford2D validation partition. Furthermore, EADO reduces trainable parameters by $67.5\%$ compared with LMO at a comparable computational cost. Future work will extend EADO to non-Cartesian sampling and volumetric reconstruction.

\section{LLM Usage Statement}
During the preparation of this work, the authors used ChatGPT to assist with language polishing, grammar checking and Latex formatting. After using these tools, the authors reviewed and edited the content as needed, independently verified the assumptions, derivations, and logical steps of all AI-assisted proofs, and take full responsibility for the correctness and integrity of the published article.
\section{Acknowledgments}
AIAR gratefully acknowledges the Tsinghua-PolyU Joint Research Initiative Fund and the Tsinghua University Dushi Program.  This work is also supported by the Yau Mathematical Sciences Center, Tsinghua University.
CWC and JAMZ are supported by the Swiss National
Science Foundation under grant number 20HW-1 220785. 
CBS acknowledges support from
the Philip Leverhulme Prize, the Royal Society Wolfson Fellowship, the EPSRC
grants EP/S026045/1, EP/T003553/1, EP/N014588/1, the Wellcome Innovator
Award RG98755 and the Alan Turing Institute. 
{
    \small
    \bibliographystyle{ieeenat_fullname}
    \bibliography{main}
}

\addtocontents{toc}{\protect\setcounter{tocdepth}{2}}

\clearpage
\onecolumn
\begin{table}[h!]
\centering
\begin{tabular}{c}
\begin{tabular}[c]{@{}c@{}} \Large  \textsc{Supplemental Material} \smallskip  \\\Large \textbf{Don’t Guess Pixels. Learn Relations! Complex Operators for MRI
Reconstruction} \end{tabular}
\end{tabular}
\end{table}

\bigskip

\begingroup
\renewcommand{\contentsname}{Supplementary Contents}
\setcounter{tocdepth}{2}
\tableofcontents
\endgroup

\bigskip


\section{Proofs of Theoretical Results}
\label{app:proofs}

\paragraph{Standing notation.}
The grid is $\mathcal G=\mathbb Z_H\times\mathbb Z_W$, $N=HW$; pixels $i,j$,
frequencies $k$; $\mathrm{i}=\sqrt{-1}$ (never an index);
$k\cdot\delta:=k_1\delta_1/H+k_2\delta_2/W$ and
$\mathcal F_{kj}=N^{-1/2}e^{-2\pi\mathrm{i}k\cdot j}$. The PSF is
$p(\delta)=\frac1N\sum_k m_k e^{2\pi\mathrm{i}k\cdot\delta}$ (the
$\operatorname{ifftshift}$ in Eq.~\eqref{eq:psf} only relabels frequencies),
so $p(0)=\rho$. Edges are $e=(i,j)$ with $j=i-\delta_e$ modulo the grid;
$U=\operatorname{diag}(u)$ with $|u_i|=1$ is unitary, and $U_{\mathcal E}$ is
the diagonal unitary multiplying edge $e$ by its source phase $u_i$.
Primed symbols denote quantities computed from the gauge-transformed data
$x\mapsto Ux$, $s_{c,i}\mapsto\overline{u_i}s_{c,i}$.

\subsection{Proof of Proposition~\ref{prop:psftail} (PSF-Tail Certificate)}
\label{supp:psftail}

\emph{Factorization and tail bound.}
A direct entry computation gives
\begin{equation}
(\mathcal F^*M\mathcal F)_{ij}
=\sum_k\overline{\mathcal F_{ki}}\,m_k\,\mathcal F_{kj}
=\frac1N\sum_k m_k\,e^{2\pi\mathrm{i}k\cdot(i-j)}
=p(i-j),
\end{equation}
so $A^*A=\sum_cS_c^*\mathcal F^*M\mathcal FS_c$ has entries $p(i-j)c_{ij}$,
and $\widetilde N=\sum_cS_c^*P_{\mathcal O}S_c$ where $P_{\mathcal O}$ is the
circulant matrix with kernel $p\,\mathbf 1_{\mathcal O}$. Hence
$A^*A-\widetilde N=\sum_cS_c^*QS_c$ with $Q$ circulant with kernel
$q=p\,(1-\mathbf 1_{\mathcal O})$. Any circulant $Q_{ij}=q(i-j)$ is
diagonalized by $\mathcal F$: using
$\frac1N\sum_ke^{2\pi\mathrm{i}k\cdot\nu}=\mathbf 1[\nu=0]$,
\begin{equation}
\big(\mathcal F^*\operatorname{diag}(\widehat q)\,\mathcal F\big)_{ij}
=\frac1N\sum_k\widehat q_k\,e^{2\pi\mathrm{i}k\cdot(i-j)}
=q(i-j),
\qquad
\widehat q_k=\sum_\delta q(\delta)e^{-2\pi\mathrm{i}k\cdot\delta},
\end{equation}
so $\|Q\|_{\mathrm{op}}=\max_k|\widehat q_k|\le\sum_\delta|q(\delta)|$ by the
triangle inequality. With $\|S_c\|_{\mathrm{op}}=\|s_c\|_\infty$,
\begin{equation}
\|A^*A-\widetilde N\|_{\mathrm{op}}
\le\|Q\|_{\mathrm{op}}\sum_c\|s_c\|_\infty^2
\le\Big(\sum_{\delta\notin\mathcal O}|p(\delta)|\Big)\sum_c\|s_c\|_\infty^2,
\end{equation}
which is~\eqref{eq:tailbound}; the sum runs over directed offsets, which is
why $\mathcal O$ must contain both orientations of each retained class.

\emph{Symmetry and class score.}
For real $m$, conjugating the definition gives
$p(-\delta)=\overline{p(\delta)}$, hence $|p(-\delta)|=|p(\delta)|$. A
retained class $\gamma=\{\delta,-\delta\}$ removes mass
$\mu(\gamma)=|\gamma|\,|p(\delta)|$ from the discarded tail, with
$|\gamma|=2$ for ordinary classes and $|\gamma|=1$ for self-inverse ones.
Since the total eligible mass depends only on the mask, the discarded tail at
fixed $K$ is minimized by the $K$ largest scores $\mu(\gamma)$; when all
eligible classes are ordinary this reduces to magnitude ranking.

\emph{Equispaced comb.}
On one axis of length $n$ with
$\widetilde m(k)=\sum_{j=0}^{n/R-1}\mathbf 1[k=k_0+Rj\!\!\pmod n]$, $R\mid n$,
\begin{equation}
p_1(\delta)
=\frac{e^{2\pi\mathrm{i}k_0\delta/n}}{n}\sum_{j=0}^{n/R-1}\omega^j,
\qquad \omega=e^{2\pi\mathrm{i}R\delta/n}.
\end{equation}
Since $\omega^{n/R}=e^{2\pi\mathrm{i}\delta}=1$, the geometric sum equals
$n/R$ if $\omega=1$ (i.e.\ $R\delta\equiv0\pmod n$) and $0$ otherwise, giving
$p_1(\delta)=\frac1R e^{2\pi\mathrm{i}k_0\delta/n}\,
\mathbf 1[\delta\in\tfrac nR\mathbb Z]$: the $R$ shifts $\ell n/R$ each carry
magnitude $1/R$, and the origin $k_0$ contributes only a phase. With full
sampling along the second axis,
$p(\delta_1,\delta_2)=p_1(\delta_1)\,\mathbf 1[\delta_2=0]$ since
$\frac1W\sum_{k_2}e^{2\pi\mathrm{i}k_2\delta_2/W}=\mathbf 1[\delta_2=0]$. The
$R-1$ nonzero shifts pair as $\ell\sim R-\ell$ under sign reversal, with
$\ell=R/2$ self-inverse when $R$ is even, giving $\lfloor R/2\rfloor$
undirected classes. \hfill$\square$

\subsection{Proof of Proposition~\ref{prop:gauge} (Gauge Covariance)}
\label{supp:gauge}

\emph{Encoding and graph.}
Since $S'_c=S_cU^*$, we have $A'=AU^*$ and $A'U=A$, hence
$A'^*A'=UA^*AU^*$ and $A'^*=UA^*$. The mask is unchanged, so $p'=p$;
$r'_i=\sum_c|s'_{c,i}|^2=r_i$; and
\begin{equation}
c'_{ij}=\sum_c u_i\overline{s_{c,i}}\,s_{c,j}\overline{u_j}
=u_i\overline{u_j}\,c_{ij},
\qquad
q'_e=u_i\overline{u_j}\,q_e .
\end{equation}
Because $|q'_e|=|q_e|$, the safeguarded denominator
$\max(|q_e|,\epsilon_g)$ is invariant, so
$\tau'_e=u_i\overline{u_j}\,\tau_e$ in all cases (including
$|\tau_e|<1$); likewise $\chi'_{ij}=\chi_{ij}$ and $\kappa'_e=\kappa_e$.
For $e=(i,j)$,
\begin{equation}
(D'Ux)_e=u_ix_i-\tau'_e\,u_jx_j=u_i\big(x_i-\tau_ex_j\big)=u_i(Dx)_e,
\end{equation}
i.e.\ $D'U=U_{\mathcal E}D$.

\emph{Pipeline.}
Since $A'^*=UA^*$ and $r'_i=r_i$, we have
$z'_i=(A'^*y)_i/\max(r'_i,10^{-8})=u_i z_i$; since $|z'_i|=|z_i|$,
the scale satisfies $s'=s$, so
$x'^0=Ux^0$ and $\bar y'=\bar y$. Inductively, if $x'^t=Ux^t$ then
$g'^t=b'-A'^*A'x'^t=U\big(b-A^*Ax^t\big)=Ug^t$. Every scalar predictor input
is invariant: magnitudes are unaffected,
$\overline{u_i}u_i=1$ removes the phase from $\overline{x'^t}g'^t$ and from
the endpoint correlation
$\overline{x'^t_i}\,\tau'_e x'^t_j=\overline{x_i^t}\,\tau_e x_j^t$,
and $\kappa_e,\zeta_e,\rho$ are unchanged. Since the network is a
deterministic function of these invariant inputs,
$\alpha'^t=\alpha^t$, $\beta'^t=\beta^t$, $w'^t=w^t$. Each carrier entry
picks up the source phase, $h'^t_e=u_ih^t_e$, and
\begin{equation}
d'^t_e=(D'x'^t)_e+\frac{0.25}{\sqrt6}\sum_{\ell=1}^6\alpha'^t_{e\ell}h'^t_{e\ell}
=u_i\,d^t_e,
\end{equation}
so $d'^t=U_{\mathcal E}d^t$ and $W'_t=W_t$.

\emph{Quadratic stage.}
Conjugation by the diagonal unitary $U_{\mathcal E}$ leaves the real diagonal
$W_t$ unchanged, so
\begin{equation}
H'_t
=UA^*AU^*+\lambda_t\,UD^*U_{\mathcal E}^*W_tU_{\mathcal E}DU^*+\eta_tI
=UH_tU^*,
\end{equation}
and the right-hand side of~\eqref{eq:system} satisfies $f'_t=Uf_t$, giving
$v'^*_t=H'^{-1}_tf'_t=Uv^*_t$; equivalently $J'_t(Uv)=J_t(v)$ term by term
($A'U=A$; $W'^{1/2}_t(D'Uv-d'^t)=U_{\mathcal E}W^{1/2}_t(Dv-d^t)$ has the
same norm; $\|Uv-Ux^t\|_2=\|v-x^t\|_2$). The Jacobi preconditioner is
invariant: $P'_t=\operatorname{diag}(UH_tU^*)=P_t$, since diagonal-unitary
conjugation preserves the diagonal.

\emph{PCG and assembly.}
Initialize $v'_0=Uv_0$ (the warm start). Then
$r'_0=f'_t-H'_tv'_0=Ur_0$, and the preconditioned residual satisfies
$z'_0=P_t^{-1}Ur_0=Uz_0$ since diagonal matrices commute. All scalar inner
products entering the PCG coefficients are invariant, since
$\langle Ua,Ub\rangle=\langle a,b\rangle$; hence the stepsizes and recurrence
coefficients coincide in both runs, and the iterate, residual,
preconditioned-residual, and search-direction updates commute with $U$ by
induction. The relative-residual stopping rule is invariant, so both runs
execute the same iterations and $x'^{t+1}=Ux^{t+1}$ in exact arithmetic.
Induction over $t=0,\ldots,T-1$ and $\widehat x'=s'x'^{(T)}=U\widehat x$
complete the proof, with $|\widehat x'|=|\widehat x|$ since $|u_i|=1$.
\hfill$\square$

\subsection{Proof of Proposition~\ref{prop:identifiability} (Joint Identifiability and Error Bound)}
\label{supp:identifiability}

Let $B=[A;\sqrt\lambda\,W^{1/2}D]$. For any $v$,
\begin{equation}
\|Bv\|_2^2=\|Av\|_2^2+\lambda\|W^{1/2}Dv\|_2^2,
\end{equation}
a sum of nonnegative terms, so
$\ker B=\ker A\cap\ker(W^{1/2}D)$. Since
$B^*B=A^*A+\lambda D^*WD\succeq0$, it is positive definite exactly when
$\ker B=\{0\}$, and injectivity of $B$ is precisely unique determination of
$x$ from the pair $(Ax,W^{1/2}Dx)$: this is~\eqref{eq:identifiability}.

For the bound, the unanchored quadratic equals $\frac12\|Bv-\beta\|_2^2$ with
$\beta=(y;\sqrt\lambda W^{1/2}d)$; under the identifiability condition its
unique minimizer is $\widehat x=B^\dagger\beta$ with $B^\dagger B=I$. Writing
$Bx^\star=\beta-\xi$ with
$\xi=(\varepsilon;\sqrt\lambda W^{1/2}\delta)$ gives
$\widehat x-x^\star=B^\dagger\xi$, and
\begin{equation}
\|\widehat x-x^\star\|_2
\le\|B^\dagger\|_{\mathrm{op}}\,\|\xi\|_2
=\frac{\sqrt{\|\varepsilon\|_2^2+\lambda\|W^{1/2}\delta\|_2^2}}
{\sigma_{\min}(B)},
\end{equation}
since the singular values of $B^\dagger$ are the reciprocals of the nonzero
singular values of $B$. This is~\eqref{eq:identerror}. \hfill$\square$

\begin{samepage}
\begin{remark}[\textbf{Scope}]
The PSF certificate bounds the normal-operator couplings omitted by truncation.
The graph penalty $D^*WD$ enters separately in the anchored solve.
PCG covariance holds in exact arithmetic. Roundoff can affect update decisions
near the residual tolerance. The identifiability condition is
acquisition- and weight-dependent and can be verified numerically via
$\lambda_{\min}(A^*A+\lambda D^*WD)$. With unit single-coil sensitivity ($c_{ij}=1$, so
$\tau_e=1$ on local edges), positive weights on the connected local-edge
graph restrict vectors in $\ker(W^{1/2}D)$ to constants before alias-edge
constraints are imposed; if all positively weighted alias edges also satisfy
$\tau_e=1$, joint identifiability is equivalent to DC being sampled, while
alias edges with $\tau_e\neq1$ can eliminate the constant null mode even when
DC is not sampled. Independently of this condition, the anchored stage is
unconditionally well-posed because $H_t\succeq\eta_tI\succ0$.
\end{remark}
\end{samepage}

\subsection{Error Bound for the Anchored Finite-PCG Stage}
\label{supp:finite-pcg}

Condition on the acquisition, the positive scale $s$ from
Eq.~\eqref{eq:normalization}, and the predictions at stage $t$. If
$y=Ax^\star+\varepsilon$, define the normalized reference
$x^\dagger=x^\star/s$, measurement error $\bar\varepsilon=\varepsilon/s$,
and edge-target error $\delta_t=d^t-Dx^\dagger$. Write the right-hand
side of Eq.~\eqref{eq:system} as $f_t$ and the residual of the computed
iterate as $r_t=f_t-H_tx^{t+1}$. Let
$\mu_t=\lambda_{\min}(H_t)\geq\eta_t>0$. Then
\begin{equation}
\label{eq:supp-finite-pcg-bound}
\left\|x^{t+1}-x^\dagger\right\|_2
\leq
\frac{\sqrt{\|\bar\varepsilon\|_2^2+
              \lambda_t\|W_t^{1/2}\delta_t\|_2^2+
              \eta_t\|x^t-x^\dagger\|_2^2}}{\sqrt{\mu_t}}
+\frac{\|r_t\|_2}{\mu_t}.
\end{equation}

\paragraph{Proof.}
Define the augmented operator and error vector
\[
B_t=\begin{bmatrix}
A\\ \sqrt{\lambda_t}\,W_t^{1/2}D\\ \sqrt{\eta_t}\,I
\end{bmatrix},
\qquad
e_t=\begin{bmatrix}
\bar\varepsilon\\ \sqrt{\lambda_t}\,W_t^{1/2}\delta_t\\
\sqrt{\eta_t}\,(x^t-x^\dagger)
\end{bmatrix}.
\]
Then $B_t^*B_t=H_t$ and
$H_t(x^{t+1}-x^\dagger)=B_t^*e_t-r_t$ by substitution. Since
$H_t$ is positive definite, this gives
$x^{t+1}-x^\dagger=B_t^\dagger e_t-H_t^{-1}r_t$.
Taking norms with $\|B_t^\dagger\|_{\mathrm{op}}=1/\sqrt{\mu_t}$ and
$\|H_t^{-1}\|_{\mathrm{op}}=1/\mu_t$ gives Eq.~\eqref{eq:supp-finite-pcg-bound}.
\hfill$\square$

The bound holds for the frozen anchored stage, including when the
unanchored joint-kernel condition fails. The first term contains the
measurement, predicted-difference and anchor errors. The second uses the
attained normal-equation residual. At the final stage the error in the
original scale is $\|\widehat{x}-x^\star\|_2=s\|x^T-x^\dagger\|_2$.
The eight-iteration budget can leave a residual above the requested
tolerance, and both the edge-target and anchor errors depend on the
reconstruction. The bound therefore gives a conditional stage-wise
estimate.

\begingroup
\graphicspath{{sec/}}
\clearpage
\section{Implementation and Evaluation Details}
\label{supp:implementation}

\paragraph{EADO implementation.}
Section~\ref{sec:supp-eado-hyperparameters} gives the architecture and training settings.
Within each stage, we hold the predicted targets and weights fixed during the linear solve.
We warm-start matrix-free, Jacobi-preconditioned conjugate gradients at $x^t$ and
recompute the final normal-equation residual. All reported EADO models use soft data fidelity.

\paragraph{IXI.}
The single-coil experiments use ten fixed $256\times256$ T2-weighted slices
per subject. The training, validation and held-out partitions contain
$401/58/116$ distinct subjects and $4{,}010/580/1{,}160$ slices, respectively.
Slices are scaled to $[0,1]$ and embedded with zero phase before Cartesian
Fourier undersampling. We trained the IXI EADO models for 100 epochs with batch size one.
After checkpoint selection on the validation split, we evaluated the controlled ablations
on the same held-out slices. The primary IXI metrics use magnitude images clipped to $[0,1]$.

\paragraph{Stanford2D.}
This multicoil study uses 71 training volumes (1,547 slices) and 18 validation volumes (490 slices). Coil images are
center-cropped or reflection-padded to $384\times384$ before forming the
retrospectively sampled k-space. Sensitivities are estimated from the 15
observed central autocalibration signal (ACS) columns by dividing each
low-resolution coil image by the corresponding root-sum-of-squares (RSS)
image, with safeguards for low signal. Input scaling uses the 99th
percentile of the ACS RSS image. Scores compare physical-scale RSS images
against the RSS reference, without clipping, using each source volume's
maximum as the PSNR and SSIM data range. Within each volume, PSNR is
computed from the mean slice MSE and SSIM is averaged across slices.
The final aggregate gives each volume equal weight. We trained the native EADO and
LMO-SENSE models for 50 epochs and selected the checkpoint with the highest mean
per-volume PSNR on the full validation set. Exact ties favor the earlier epoch.

\paragraph{Mask and baseline conventions.}
For each fixed-weight comparison, all methods receive the same realized evaluation mask.
Acceleration is the grid size divided by the number of sampled Fourier
coefficients, including the ACS region. The native IXI masks reach the target
rate except the nominal \(4\times\) radial-grid mask, which samples
19{,}268 of 65{,}536 coefficients (\(R=3.4013\)). Radial-grid masks
are rasterized on the Cartesian Fourier grid. The native comparisons use mask-specific checkpoints. The Stanford2D unseen-mask study in
Sec.~\ref{sec:supp-additional-stanford} fixes the Random-trained checkpoint of each model
across target masks. The Stanford2D LMO adaptation uses explicit SENSE data consistency
and per-image BatchNorm at inference.

\paragraph{Inference operation counts.}
The main comparison estimates one full forward pass at batch size one on
a $384\times384$ grid with 16 acquired coils, a common Cartesian $8\times$
mask and supplied sensitivity maps. The convention counts two FLOPs per
multiply--accumulate and $5N\log_2 N$ per complex-equivalent FFT of size
$N$, plus explicit normalization. It includes learned layers, FFTs,
solver vector operations and the EADO graph operations. LMO-SENSE uses
six cascades with ten CG iterations per cascade. The resulting estimates
are 744.99 GFLOPs for EADO and 738.15 GFLOPs for LMO-SENSE. The benchmark
boundary starts from prepared k-space and includes conversion to the output
RSS image. Calibration, file I/O and training fall outside this boundary.
The values describe arithmetic work under the specified convention.

\clearpage

\section{Additional Qualitative Results}
\label{sec:supp-qualitative}

Figures~\ref{fig:supp-qual-ixi}, \ref{fig:supp-qual-stanford}, \ref{fig:supp-qual-ixi-x8}, and~\ref{fig:supp-qual-stanford-x4} compare Random, Radial, and Equispaced reconstructions. Each example uses the same input and crop locations across methods. The last two columns show absolute errors against the reference on a shared color scale. Brighter colors indicate larger errors. 

\begin{figure}[!htbp]
    \centering
    \includegraphics[width=0.90\linewidth]{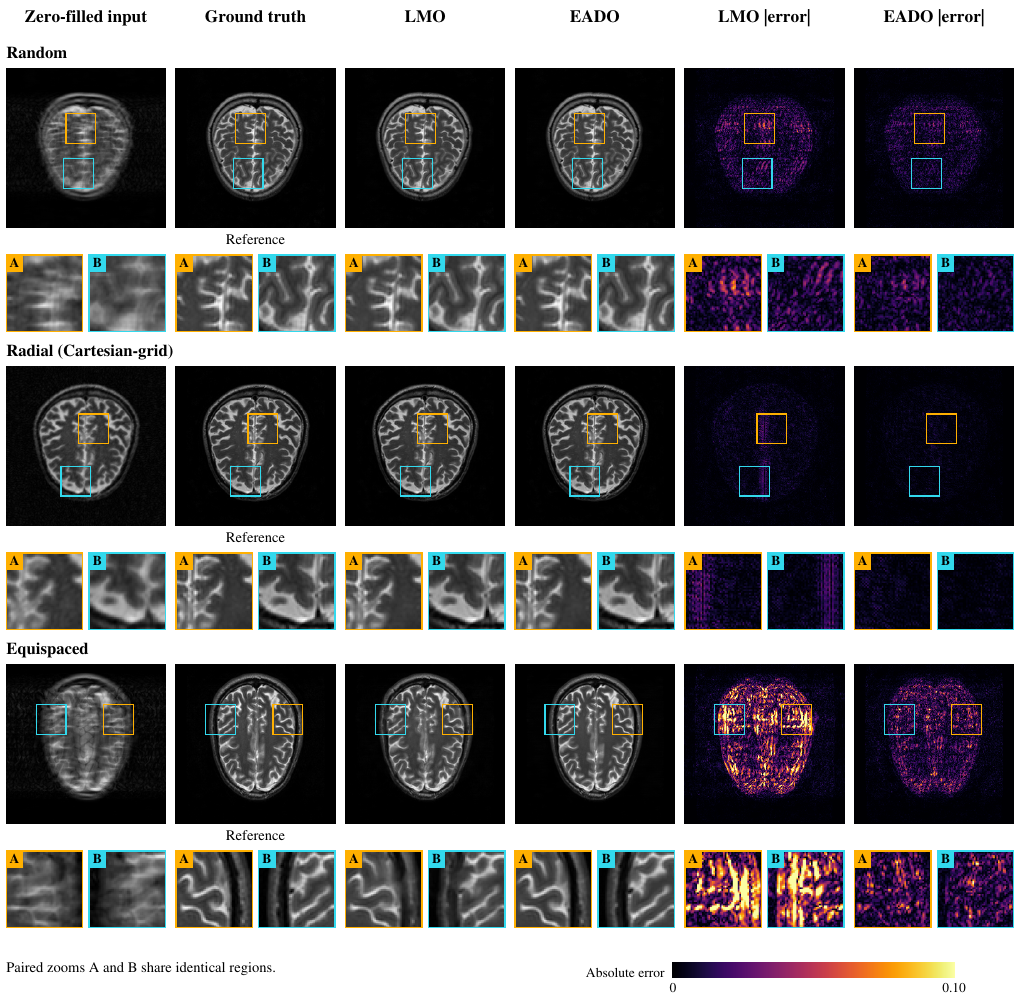}
    \caption{\textbf{Structured residuals in nominal $\times4$ IXI reconstructions.} The zero-filled input, reference, and two reconstructions are displayed alongside absolute-error maps. Yellow and cyan boxes mark the matched zooms A and B. In the Equispaced row, the bright residual pattern in the LMO zooms is weaker in the corresponding EADO zooms.}
    \label{fig:supp-qual-ixi}
\end{figure}

\clearpage
\noindent\textbf{Stanford2D comparison.}
We compare the native mask-specific $\times8$ checkpoints on slices from the Stanford2D validation cohort used for checkpoint selection. LMO-SENSE uses per-image BatchNorm at inference.

\begin{figure}[!htbp]
    \centering
    \includegraphics[width=0.90\linewidth]{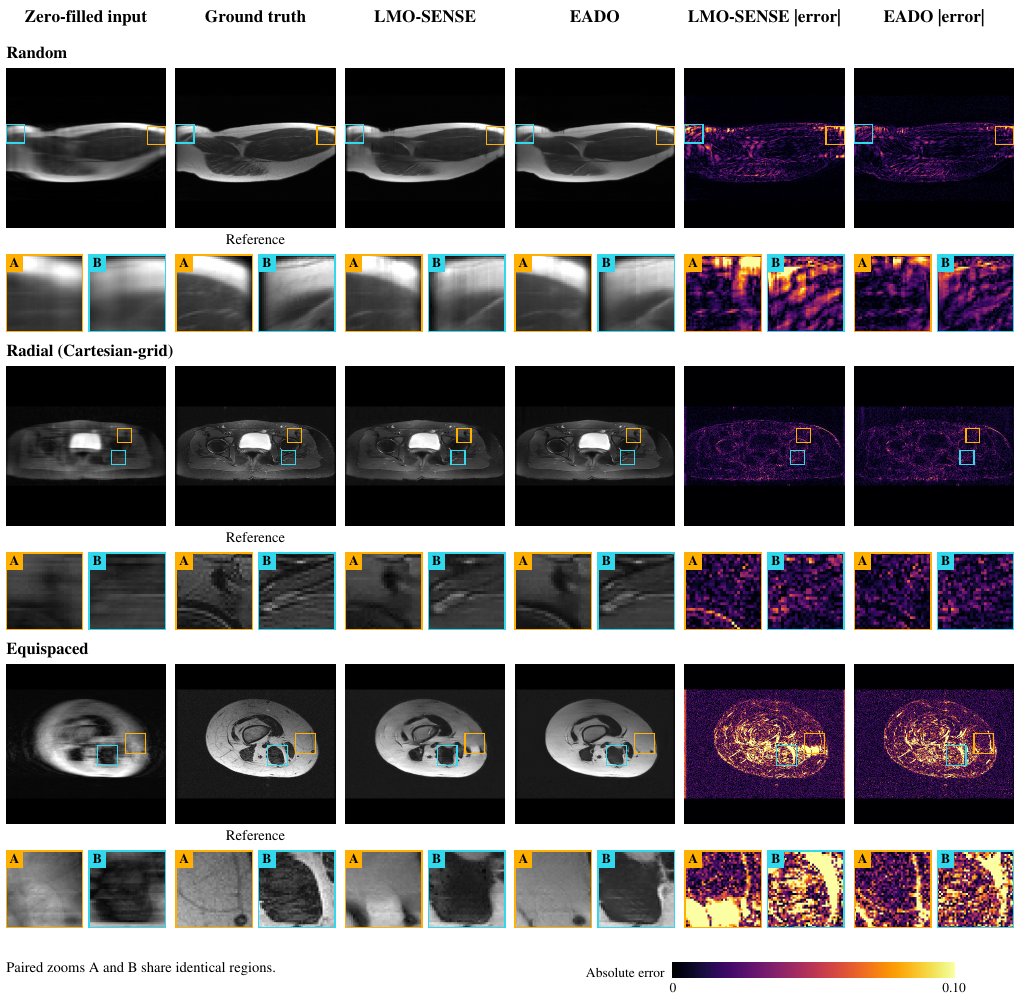}
    \caption{\textbf{Stanford2D $\times8$ error maps localize differences near tissue boundaries.} The Random and Equispaced rows show weaker bright residuals around the marked boundaries with EADO. The Radial pair is visually closer. Insets A and B show the same locations across methods. These slices were selected to show positive differences. Across the validation cohort, LMO-SENSE leads on mean Radial PSNR and SSIM and mean Random SSIM.}
    \label{fig:supp-qual-stanford}
\end{figure}

\clearpage
\begin{figure}[!htbp]
    \centering
    \includegraphics[width=0.90\linewidth]{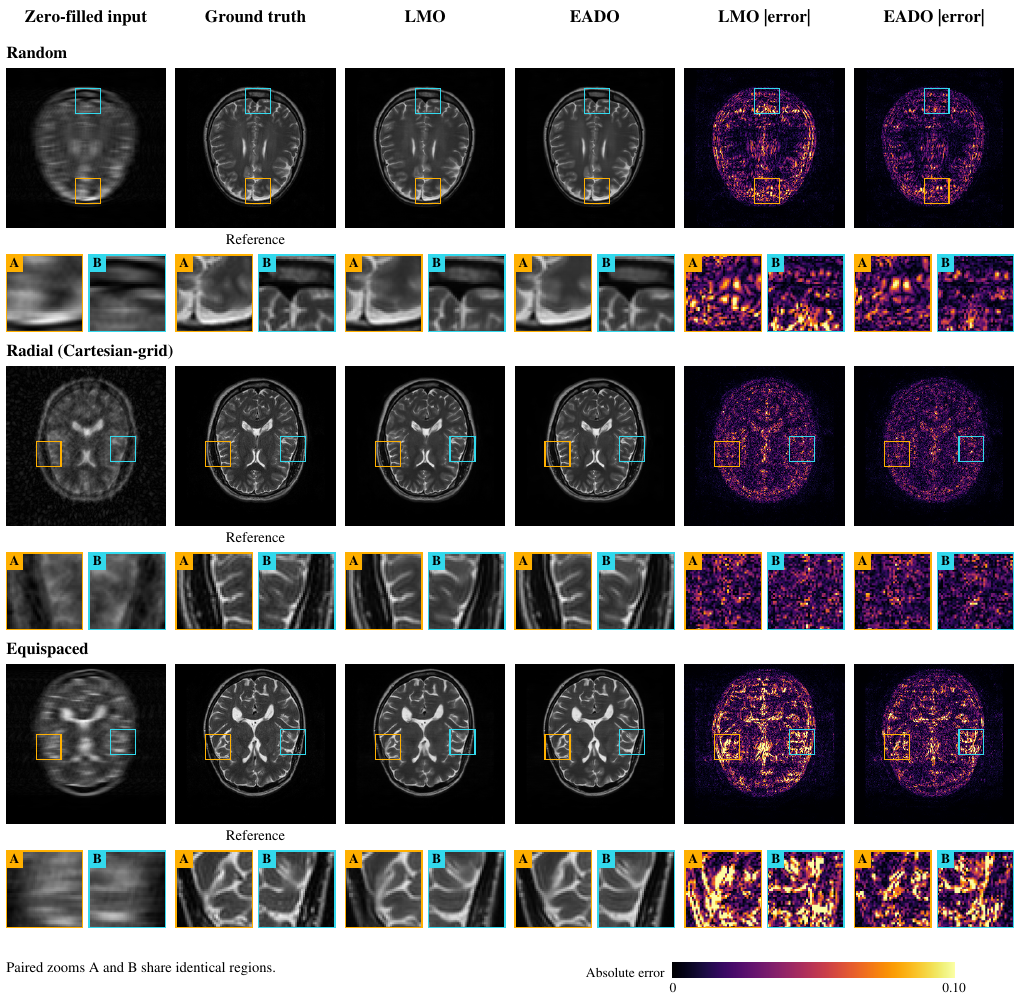}
    \caption{\textbf{Matched IXI $\times8$ examples show weaker residuals with EADO.} The zooms reveal brain contours blurred in the zero-filled inputs. The Equispaced row retains the brightest residuals, with a darker EADO error map in both marked regions. Each row shows the held-out slice nearest the median paired PSNR difference for its mask.}
    \label{fig:supp-qual-ixi-x8}
\end{figure}

\clearpage
\begin{figure}[!htbp]
    \centering
    \includegraphics[width=0.90\linewidth]{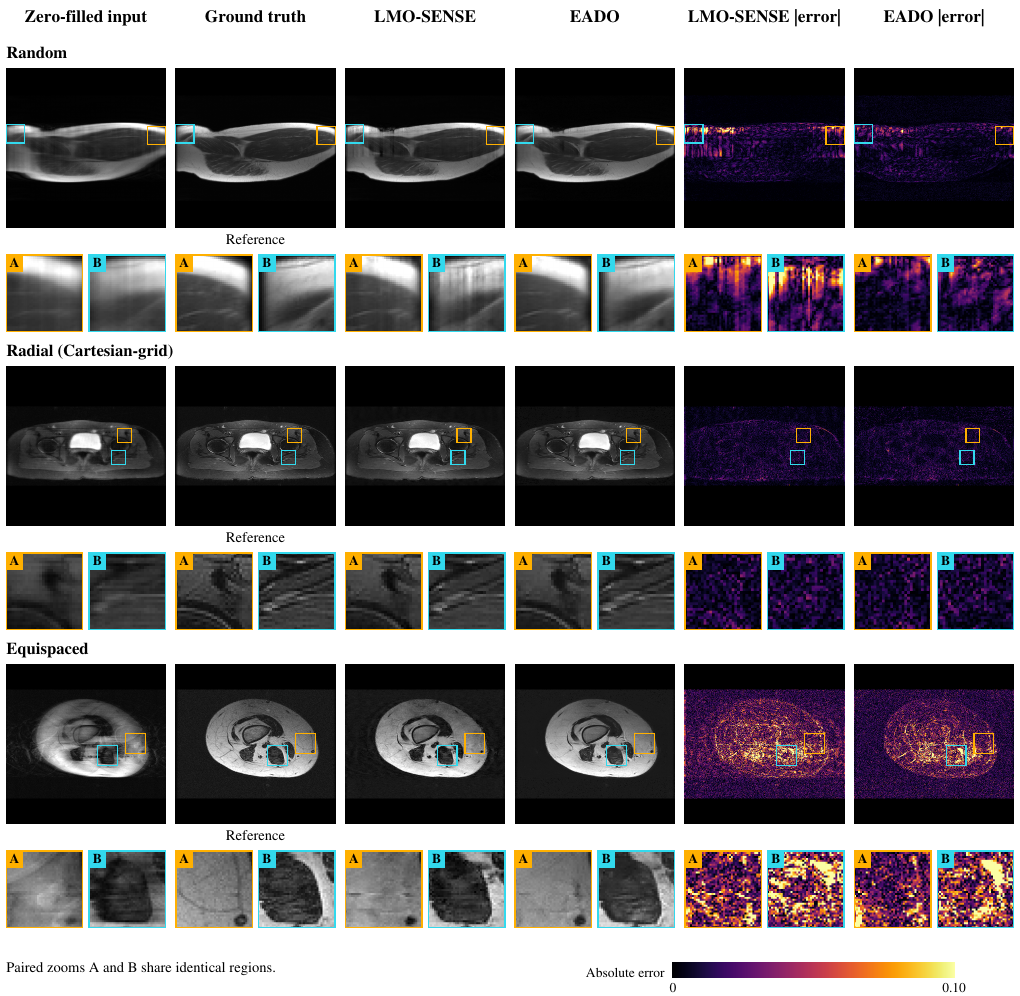}
    \caption{\textbf{The same Stanford2D cases show mixed visual differences at $\times4$.} In the Random row, EADO reduces bright residuals in the marked regions. The Radial reconstructions and their error maps look similar. Visible residuals persist for both methods in the Equispaced row. We retained the cases and crops from Fig.~\ref{fig:supp-qual-stanford} before inspecting the $\times4$ reconstructions.}
    \label{fig:supp-qual-stanford-x4}
\end{figure}

\clearpage

\clearpage
\endgroup

\begingroup
\graphicspath{{sec/}}
\section{Stanford2D Transfer to Unseen Masks}
\label{sec:supp-additional-stanford}
We compare Random-trained $\times8$ EADO and LMO (SENSE-adapted) using absolute scores and volume-level paired differences across all 12 target settings. For each method, we selected on the original validation task and held fixed across target masks. LMO uses per-image BatchNorm with running statistics disabled.

\paragraph{Evaluation protocol.}
Every target setting uses the same three saved masks for both methods, with exact acceleration $R\in\{4,6,8,12\}$ and 15 central ACS columns. Metrics average three mask realizations within each volume, then weight the 18 volumes equally. We evaluate all 490 slices from the 18-volume checkpoint-selection validation cohort. PSNR and SSIM use unclipped physical RSS and the source full-volume maximum. Radial-shaped masks are Cartesian-grid approximations.

\begin{table}[!ht]
\centering
\caption{\textbf{Random-trained $\times8$ EADO raises mean PSNR and SSIM across all 12 target settings.} Scores give equal weight to each validation volume. The last column reports EADO minus LMO with a paired-volume pointwise 95\% interval. Positive $\Delta$PSNR favors EADO.}
\label{tab:supp-stanford-x8-absolute}
\begingroup\fontsize{9}{10.5}\selectfont
\setlength{\tabcolsep}{3pt}
\renewcommand{\arraystretch}{1.07}
\begin{tabular*}{\linewidth}{@{\extracolsep{\fill}}lrccccc@{}}
\toprule
 & & \multicolumn{2}{c}{PSNR (dB) $\uparrow$} & \multicolumn{2}{c}{SSIM $\uparrow$} & \\
\cmidrule(lr){3-4}\cmidrule(lr){5-6}
Target pattern & $R$ & LMO & EADO & LMO & EADO & $\Delta$PSNR $[95\%\ \mathrm{CI}]$ \\
\midrule
Random & 4 & 34.568 & 35.639 & 0.8091 & 0.8841 & $ +1.071\;[+0.613,+1.432]$ \\
 & 6 & 34.569 & 35.847 & 0.8320 & 0.9021 & $ +1.277\;[+0.995,+1.557]$ \\
 & 8 & 33.851 & 35.195 & 0.8249 & 0.8987 & $ +1.344\;[+1.062,+1.633]$ \\
 & 12 & 32.010 & 32.608 & 0.7993 & 0.8604 & $ +0.599\;[+0.415,+0.784]$ \\
\addlinespace[2pt]
Equispaced & 4 & 35.448 & 36.109 & 0.8097 & 0.8870 & $ +0.661\;[+0.162,+1.070]$ \\
 & 6 & 34.455 & 34.911 & 0.8231 & 0.8940 & $ +0.456\;[-0.017,+0.919]$ \\
 & 8 & 33.938 & 34.131 & 0.8379 & 0.8880 & $ +0.192\;[-0.188,+0.567]$ \\
 & 12 & 32.666 & 32.673 & 0.8479 & 0.8655 & $ +0.007\;[-0.199,+0.206]$ \\
\addlinespace[2pt]
Radial grid & 4 & 36.953 & 37.634 & 0.8306 & 0.8948 & $ +0.681\;[+0.057,+1.165]$ \\
 & 6 & 36.965 & 37.516 & 0.8628 & 0.9066 & $ +0.551\;[+0.064,+1.014]$ \\
 & 8 & 36.521 & 36.942 & 0.8738 & 0.9042 & $ +0.421\;[+0.063,+0.791]$ \\
 & 12 & 34.917 & 35.318 & 0.8547 & 0.8894 & $ +0.401\;[+0.059,+0.733]$ \\
\bottomrule
\end{tabular*}
\endgroup
\end{table}

\paragraph{Mean gains and uncertainty.}
Random-trained $\times8$ EADO has higher mean PSNR and SSIM in all 12 target settings (Table~\ref{tab:supp-stanford-x8-absolute}). The pointwise 95\% intervals favor EADO in 9 of 12 PSNR comparisons and all 12 SSIM comparisons. The largest PSNR gain is $1.344$\,dB at Random $\times8$. At Equispaced $\times12$, the $0.0066$\,dB difference is inconclusive because its interval $[-0.1987,+0.2061]$ crosses zero.

We compute pointwise 95\% intervals from 2,000 paired-volume percentile resamples (seed 20261111), after averaging the three masks within each volume. The intervals are conditional on the fixed masks and single-seed checkpoints, without multiplicity correction.

\clearpage
\paragraph{Volume-level variation.}
Across the 12 target settings, EADO improves PSNR in 12--18 of the 18 validation volumes and SSIM in 15--18 volumes (Fig.~\ref{fig:supp-stanford-volume-consistency}). Each paired difference averages the same three mask realizations within one volume. The plot includes all 18 volumes per setting, including those that favor LMO.

\begin{figure}[!ht]
\centering
\includegraphics[width=\linewidth]{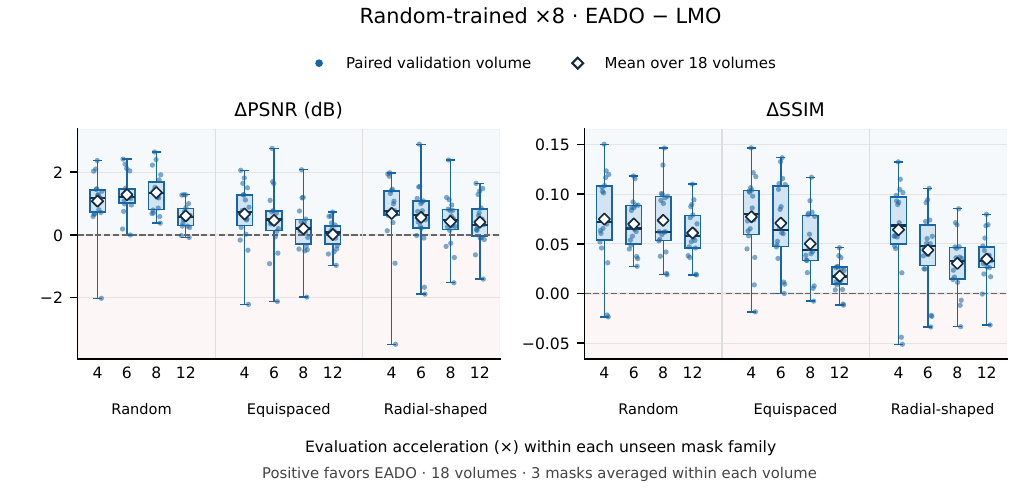}
\caption{\textbf{Most Stanford2D volumes gain PSNR and most gain SSIM.} Panels show EADO minus LMO for Random-trained $\times8$ checkpoints. Each point represents one validation volume averaged over three shared masks. Boxes show the interquartile range and median, whiskers span the full range, and diamonds mark the mean. The dashed line marks zero.}
\label{fig:supp-stanford-volume-consistency}
\end{figure}

\clearpage
\endgroup

\section{EADO Model and Training Hyperparameters}
\label{sec:supp-eado-hyperparameters}
For each dataset, we train a separate EADO model for each combination of nominal acceleration ($\times4$ or $\times8$) and sampling pattern (Random, Equispaced, or Radial grid). All 12 runs use the same architecture and start from fresh weights with seed 24.

\paragraph{Shared architecture and graph.}
EADO has eight reconstruction stages and one shared difference predictor, for $656{,}621$ trainable parameters in total. The scalar context U-Net uses 48 channels at full resolution and 96 channels at the coarser resolutions. Its convolutional blocks use GroupNorm with eight groups and SiLU activations. The edge head has two $1\times1$ hidden layers of width 48 with SiLU and a 13-channel output that supplies one weight logit and six complex coefficients. We use activation checkpointing for the edge head during training. The graph uses four local offset representatives $(1,0)$, $(0,1)$, $(1,1)$, and $(1,-1)$, together with up to six additional alias offsets selected by sampling PSF magnitude.

\paragraph{Data consistency and solver.}
Each stage uses soft data fidelity and warm-starts Jacobi-preconditioned conjugate gradients from the current reconstruction. The solver runs eight iterations with updates gated by a relative-residual tolerance of $10^{-6}$. We learn the graph and anchor strengths separately at each stage. For raw scalars $a_t$ and $b_t$ initialized to $-1$ and $-2$, respectively, we set $\lambda_t=0.01+0.99\,\sigma(a_t)$ and $\eta_t=0.005+0.195\,\sigma(b_t)$, where $\sigma$ is the logistic sigmoid.

\begin{table}[!ht]
\centering
\caption{\textbf{EADO training hyperparameters.} The base learning rate is the peak reached at the end of warmup. The cosine schedule ends at $5\%$ of that rate.}
\label{tab:supp-eado-hyperparameters}
\begin{tabular*}{\linewidth}{@{\extracolsep{\fill}}lll@{}}
\toprule
Setting & IXI & Stanford2D \\
\midrule
Reconstruction grid & $256\times256$ & $384\times384$ \\
Training epochs & 100 & 50 \\
Batch size (slices) & 1 & 1 \\
Optimizer & AdamW & AdamW \\
Base learning rate & $4\times10^{-3}$ & $1\times10^{-4}$ \\
Final scheduled learning rate & $2\times10^{-4}$ & $5\times10^{-6}$ \\
Weight decay & $10^{-5}$ & $10^{-5}$ \\
Global gradient norm clip & 5 & 5 \\
Charbonnier $\epsilon$ & $10^{-3}$ & $10^{-3}$ \\
Supervised output & Image magnitude & Coil RSS magnitude \\
\bottomrule
\end{tabular*}
\end{table}


\end{document}